\documentclass[aoas,nameyear,dvips]{arximspdf}
\usepackage{dcolumn}
\usepackage{graphicx}

\doi{10.1214/09-AOAS284}
\volume{4}
\issue{1}
\pubyear{2010}
\firstpage{340}
\lastpage{365}

\makeatletter
\newcolumntype{d}[1]{D{.}{.}{#1}}
\newtheorem{theorem}{Theorem}
\newtheorem{assumption}{Assumption}
\makeatother

\begin{document}
\begin{frontmatter}

\title{A multivariate adaptive stochastic search method for
dimensionality reduction in classification}
\runtitle{Multivariate Adaptive Stochastic Search}

\begin{aug}
\author[A]{\fnms{Tian Siva} \snm{Tian}\ead[label=e1]{ttian@mail.uh.edu}\corref{m2}},
\author[B]{\fnms{Gareth M.} \snm{James}\ead[label=e2]{gareth@usc.edu}} \and
\author[C]{\fnms{Rand R.} \snm{Wilcox}\ead[label=e3]{rwilcox@usc.edu}}
\runauthor{T. S. Tian, G. M. James and R. R. Wilcox}
\affiliation{University of Houston, University of Southern California
and University~of~Southern California}
\address[A]{T. S. Tian\\ Department of Psychology\\ University of
Houston\\ Houston, Texas 77204\\USA\\ \printead{e1}} 
\address[B]{G. M. James\\ Department of Information and\\\quad
Operations Management\\ University of Southern California\\ Los
Angeles, California 90089\\USA\\ \printead{e2}}
\address[C]{R. R. Wilcox\\ Department of Psychology\\ University of
Southern California\\ Los Angeles, California 90089\\USA\\ \printead{e3}}
\end{aug}

\received{\smonth{11} \syear{2008}}
\revised{\smonth{7} \syear{2009}}

%
\begin{abstract}
High-dimensional classification has become an increasingly important
problem. In this paper we propose a ``Multivariate Adaptive Stochastic
Search'' (MASS) approach which first reduces the dimension of the data
space and then applies a standard classification method to the reduced
space. One key advantage of MASS is that it automatically adjusts to
mimic variable selection type methods, such as the Lasso, variable
combination methods, such as PCA, or methods that combine these two
approaches. The adaptivity of MASS allows it to perform well in
situations where pure variable selection or variable combination
methods fail. Another major advantage of our approach is that MASS can
accurately project the data into very low-dimensional non-linear, as
well as linear, spaces. MASS uses a stochastic search algorithm to
select a handful of optimal projection directions from a large number
of random directions in each iteration. We provide some theoretical
justification for MASS and demonstrate its strengths on an extensive
range of simulation studies and real world data sets by comparing it to
many classical and modern classification methods.

\end{abstract}

%

\end{frontmatter}
%

\section{Introduction}

An increasingly important topic is the classification of observations into
two or more predefined groups when the number of predictors, $d$, is larger
than the number of observations, $n$. For example, one may need to
identify at what time a subject is performing a specific task based on
hundreds of thousands of brain voxel values in a functional Magnetic
Resonance Imaging (fMRI) study, where changes in blood flow and blood
oxygenation are measured when brain neurons are activated. In
particular, the data set that motivated the development of the
methodology in this paper was an fMRI study based on vision research.
We wish to predict, at a given time, whether a subject is conducting a
task or resting (baseline), based on the activity level of the observed
voxels in the subject's brain. This is a difficult task because the
number of voxels, $d$, is much higher than the number of observations,
$n$. In this case, many conventional classification methods, such as
Fisher's discriminant rule, are not applicable since $n<d$. Other
methods, such as classification trees, $k$-nearest neighbors and
logistic discriminant analysis, do not explicitly require $n>d,$ but in
practice provide poor classification accuracy in this situation.

A common solution is to first reduce the data into a lower, $p\ll d$
dimensional space and then perform classification on the transformed
data. For example, \citet{FL08} provide significant theoretical and
empirical evidence for the power of such an approach. Generally, the
dimension reduction is performed using a linear transformation of the form
%
\begin{equation} \label{eq:linear}
\mathbf{Z}=\mathbf{X}\mathbf{A},
\end{equation}
where $\mathbf{X}$ is an $n$-by-$d$ data matrix and $\mathbf{A}$ is a
$d$-by-$p$ transformation matrix which projects $\mathbf{X}$ onto a
$p$-dimensional subspace $\mathbf{Z}~(p<d)$. However, there are many
possible approaches to choosing $\mathbf{A}$. We divide the methods into
supervised versus unsupervised and variable selection versus variable
combination.

Variable selection techniques select a subset of relevant variables that
have good predictive power, thus obtaining a subset of informative
variables from a set of more complex variables. In
this setting, $\mathbf{A}$ is some row permutation of the identity
matrix and the
zero matrix, that is, $\mathbf{A}=\operatorname{perm}([\mathbf{I}_p,\mathbf{0}_{(d-p) \times p}]^T)$. If we define sparsity as the
fraction of zero elements in a given matrix, then $\mathbf{A}$ would
be considered sparse because most of its components are zeros. Many
variable selection methods have been proposed and widely used in
numerous areas. A great deal of attention is paid to the $L_1$
penalized least squares estimator [i.e., the Lasso \citet{T96}, \citet{EHJT04}]. Other methods include SCAD \citet{FL01}, nearest
shrunken centroids \citet{THNC02}, the Elastic Net \citet{ZH05},
Dantzig selector \citet{CT07}, VISA \citet{RJ08}, FLASH \citet{RJ09}
and Bayesian methods of variable selection \citet{MB88}, \citeauthor{GM93} (\citeyear{GM93,GM97}).
These methods all use supervised learning, where the response and
predictors are both utilized to obtain the subset of variables. In
addition, because these methods always select a subset of the original
variables, they provide highly interpretable results. However, because
of the sparsity of $\mathbf{A}$, for a given $p$, variable selection
methods are less efficient at compressing the observed data, $\mathbf
{X}$. For example, they may discard potentially valuable variables
which are not
predictive individually but provide significant improvement in
conjunction with others.

In comparison, variable combination methods utilize a dense $\mathbf
{A}$ which combines correlated variables and hence does well on
multicollinearities which often occur in high-dimensional data.
Probably the most commonly applied method in this category is principal
component analysis (PCA). Using this approach, $\mathbf{A}$ becomes
the first $p$ eigenvectors of $\mathbf{X}$, and $\mathbf{Z}$ is the
associated PCA scores. PCA can deal with an ultra large data scale and
produces the most efficient representation of $\mathbf{X}$ using $p$
dimensions. However, PCA is an unsupervised learning technique which
does not make use of the response variable to construct $\mathbf{A}$.
It is well known that the dimensions that best explain $\mathbf{X}$
will not necessarily be the best dimensions for predicting the response
$Y$. Other variable combination methods include partial least squares
regression and multidimensional scaling. All these approaches yield
linear combinations of variables which makes interpretation more difficult.

In this paper we propose a new supervised learning method which we call
``Multivariate Adaptive Stochastic Search'' (MASS). Our approach works
by projecting a high dimensional data set into a lower dimensional space
and then applying a classifier to the projected data. However, MASS has
two key advantages over these other methods. First, when using a linear
projection, such as given by (\ref{eq:linear}), MASS uses a stochastic
search process that is capable of automatically adapting the sparsity of
$\mathbf{A}$ to generate optimal prediction accuracy. Hence, in
situations where
a subset of the original variables provides a good fit, MASS will utilize
a sparse model, while in situations where linear combinations of the
variables work better, MASS will produce a denser model. MASS has the same
advantage as other supervised methods in which $\mathbf{A}$ can be designed
specifically to provide the best level of prediction accuracy. However, it
has the flexibility to adapt the sparsity of $\mathbf{A}$ so as to
gain the
benefits of both the variable selection and variable combination
methods. The second major advantage of MASS is that, with only a small
adaption to the standard fitting procedure, it can project the original
data into a nonlinear space. This generalization of (\ref{eq:linear})
potentially allows for a very accurate projection into a low-dimensional
space with little additional effort.

MASS starts by generating a large set of prospective columns for
$\mathbf{A}$
with a given sparsity level. A variable selection technique such as the
Lasso is then applied to select a candidate subset of ``good'' columns or
directions. Then, a new set of columns with a new sparsity level
are produced as candidates. The variable selection method must choose among
the current set of good columns and the new candidates. Over time the same
best columns are picked at each iteration and the process converges. At
each step in the algorithm, the sparsity level of the new prospective
columns is adjusted according to the sparsity level of the previously
chosen columns. We show through extensive simulations and real world
examples that MASS is highly robust in that it generally provides
comparable performance to variable selection and variable combination
approaches in situations that favor each of these methods. However, MASS
can still perform well in situations where one or another of these
approaches fails.

This paper is organized as follows. In Section~\ref{sec:2} we present
the basic MASS methodology. After outlining the linear fitting
algorithm, we show how this can easily be extended to the nonlinear
generalization. Furthermore, we provide some theoretical motivation for
MASS and discuss a preliminary data reduction which can be implemented
before applying MASS. In Section~\ref{sec:3} we demonstrate the
performance of MASS on an fMRI study and a gene microarray study. In
Section~\ref{sec:4} we further study the performance of MASS on
different scenarios by comparing MASS with several other modern
potential classification techniques, such as \mbox{$k$-nearest} neighbors,
support vector machines, random forests and neural networks, in
extensive simulation studies. We briefly investigate some issues in
implementing MASS, such as solution variability, computational cost and
the problem of overfitting, in Section~\ref{sec:5}. A brief discussion
summarizes the paper in Section~\ref{sec:6}.

\section{Projection selection with MASS} \label{sec:2}

\subsection{General ideas and motivations}

Given predictors, $\mathbf{x}_i \in\mathbb{R}^d$, and corresponding categorical
responses, $y_i$, we model the relationship between $y_i$ and $\mathbf{x}_i$
as
%
\begin{eqnarray}
\label{yeqn}y_i|\mathbf{x}_i &\sim& g(y_i|\mathbf{z}_i),\qquad   i=1,\dots
,n,\\
\label{zeqn} z_{i,j}&=&f_j(\mathbf{x}_i),\qquad j=1,\dots,p,
\end{eqnarray}
where
$\mathbf{z}_i=(z_{i,1},\dots,z_{i,p})$ for some $p\ll d$. Our general approach
is to estimate the $f_j$'s, project the data into a lower
$p$-dimensional sub-space, $\mathbf{z}_i$, and then apply a standard
classification method to fit (\ref{yeqn}).

To make this problem tractable, we further assume that $f_j$ has an
additive structure,
$f_j(\mathbf{x}_i)=\sum^d_{k=1}f_{j,k}(x_{i,k})$, and, hence, (\ref
{zeqn}) becomes
%
\begin{equation} \label{eq:general}
z_{i,j}=\sum^d_{k=1}f_{j,k}(x_{i,k}).
\end{equation}
For any given $z_{i,j}$ and $x_{i,k}$'s, there are many functions,
$f_{j,k}$, that satisfy
(\ref{eq:general}). However, to solve Equation (\ref{eq:general}), we
constrain the flexibility of these functions by imposing the
constraints, $f_{j,k}(0)=0$, and
%
\begin{equation}
\label{smooth.constraint}
\int f_{j,k}''(x)^2\,dx \le\lambda,\qquad    j=1,\ldots,p,
k=1,\ldots, d,  \lambda\le0.
\end{equation}
It should be noted that Equation (\ref{smooth.constraint}) can
trivially be made to hold by rescaling $f_{j,k}$. To prevent this
occurring, we impose a further constraint on the first derivative of
the $f_{j,k}$'s, when $f_{j,k}$'s are nonlinear; details are proved in
Appendix~\ref{app:B}. Using equation (\ref{smooth.constraint}), small
values of $\lambda$
constrain $f_{j,k}$ to be close to a linear function. In particular,
setting $\lambda=0$ implies $f_{j,k}(x)=a_{j,k}x$, in which case (\ref
{zeqn}) reduces to the linear projection given by (\ref{eq:linear}).
We first describe the linear MASS approach for fitting (\ref{yeqn})
and (\ref{eq:general}) subject to $\lambda=0$ and then in
Section \ref{generalsec} show how the procedure can easily be extended to the more
general nonlinear case when $\lambda>0$.

In the linear situation fitting our model given by (\ref{yeqn}) through
(\ref{eq:general}) requires choosing the $f_{j,k}$'s, or equivalently
estimating $\mathbf{A}$ in (\ref{eq:linear}), and also selecting a
classifier to
apply to the lower dimensional data. We place most attention on the former
problem because there are many classification techniques that have been
demonstrated to perform well on low-to-medium dimensional data. A more
difficult question involves the best way to produce the lower dimensional
data. Hence, we assume that one of these classification methods has been
chosen and concentrate our attention on the choice of $\mathbf{A}$.
This choice
can be formulated as the following optimization problem:
%
\begin{eqnarray} \label{opt}
&&\mathbf{A}=\arg\min_{\mathbf{A}}E_{\mathbf{X},Y}[e(\mathcal
{M}_{\mathbf{A}}(\mathbf{X}),Y)] \\
&&\mbox{subject to }\quad  \Vert \mathbf{a}_j\Vert =1, \nonumber
\end{eqnarray}
where $\mathcal{M}_{\mathbf{A}}$ is the classification method applied to
the lower dimensional data, ${\mathbf{X}}$~and~$Y$ are the predictors and
response variables, and $e$ is a loss function resulting from using
$\mathcal{M}_{\mathbf{A}}$ to predict $Y$. The constraint is that
each column of $\mathbf{A}$ should be norm 1. A common choice for $e$
is the $0$--$1$ loss function which results in minimizing the misclassification
rate (MCR). Since $\mathbf{A}$ is high dimensional, solving (\ref
{opt}) is in
general a difficult problem.

There are several possible approaches to optimize (\ref{opt}). One option
is to assume that $\mathbf{A}$ is a sparse $0,1$ matrix and only
attempt to
estimate the locations of its non-zero elements. This is the approach
taken by variable selection methods. Another option is to assume
$\mathbf{A}$ is
dense but, instead of choosing $\mathbf{A}$ to optimize (\ref{opt}),
select a
matrix which provides a good representation for $\mathbf{X}$. This is the
approach taken by~PCA. One then hopes that the PCA
solution will be close to that of (\ref{opt}).

Instead of making restrictive assumptions about $\mathbf{A}$, as with the
variable selection approach, or failing to directly optimize (\ref{opt}),
as with the PCA approach, we attempt to directly fit (\ref{opt}) without
placing restrictions on the form of $\mathbf{A}$. In this type of high
dimensional nonlinear
optimization problem, stochastic search methods, such as genetic
algorithms and simulated annealing, have been
shown to provide superior results over more traditional deterministic
procedures because they are often able to more effectively search large
parameter spaces, can be used for any class of objective functions and
yield an asymptotic guarantee of convergence \citet{G03}, \citet{LK05}.
We explore a stochastic search process and demonstrate that it is highly
effective at searching the parameter space and generally requires significantly
fewer iterations than other possible stochastic approaches \citet{TWJ09}.

\subsection{The MASS method}\label{sec2.2}

The linear MASS procedure works by successively generating a large number,
$L$, of potential random directions, that is, $\mathbf{a}_j$'s. We
then use Equation
(\ref{eq:linear}) to compute the corresponding $L$ dimensional data space,
$\mathbf{z}_1, \mathbf{z}_2, \ldots, \mathbf{z}_L$, and use a
variable selection method to
select a ``good'' subset of these directions to form an initial
estimate, $\mathbf{A}^*$. The sparsity level of this $\mathbf{A}^*$
is examined and a new random set of potential columns is generated with
the same average
sparsity as the current $\mathbf{A}^*$. The procedure iterates in this fashion
for a fixed number of steps. Formally, the MASS procedure consists of the
following steps:
\begin{enumerate}[Step 5.]
\item[Step 1.] Randomly generate the initial $d$-by-$L$ transformation matrix
$\mathbf{A}^{*(0)}$ $(p<L<d)$ with an expected sparsity of $\bar{\xi}^{(0)}$.

\item[Step 2.] At the $l$th iteration, use Equation (\ref{eq:linear})
and $\mathbf{A}^{*(l)}$ to
obtain a preliminary reduced data space $\mathbf{Z}^{*(l)}$. Evaluate
each variable
of $\mathbf{Z}^{*(l)}$ by fitting a model $Y\sim\mathbf{Z}$, and
select the $p$ most ``important'' variables in terms of the model.

\item[Step 3.] Keep the corresponding $p$ columns of $\mathbf
{A}^{*(l)}$ and calculate the
average sparsity, $\bar{\xi}^{(l)}$, for these columns.

\item[Step 4.] Generate $L-p$ new columns with an average sparsity of
$\bar{\xi}^{(l)}$. Join these columns with the $p$ columns selected
in Step
3 to form a new\vspace*{1pt} transformation matrix $\mathbf{A}^{*(l+1)}$.

\item[Step 5.] Return to Step 2 for a fixed number of iterations.
\end{enumerate}

Implementing this approach requires the choice of a variable selection
procedure for Step 2. Potentially any of a large range of standard
methods can be
chosen. We discuss various options in Section~\ref{subsec:imp}.

A more crucial part of implementing MASS is the mechanism for
generating\vspace*{1pt} the potential columns
for $\mathbf{A}^*$. We define the current level of sparsity, $\bar
{\xi}^{(l)}$,
as the fraction of zero elements in $\mathbf{A}^{(l)}$. Then at each
iteration of
MASS we generate new potential columns with the same average sparsity
level as the columns selected in the previous step. This allows
$\bar{\xi}^{(l)}$ to automatically adjust to the data set. The idea is
that a data set requiring high $\bar{\xi}^{(l)}$ will tend to result in
sparser columns being selected and the current $\mathbf{A}$ will
become sparser
at each iteration. Alternatively, a data set that requires a denser
$\mathbf{A}$
will select dense columns at each iteration.
While at each step of the stochastic search the overall average
sparsity is restricted to equal $\bar{\xi}^{(l)}$, we desire some variability
in the sparsity levels so that MASS is able to select out columns with
higher or lower sparsity and hence adjust $\bar{\xi}^{(l)}$ for the next
iteration. To achieve this goal, we allow for different average sparsity
levels between columns.

In particular, we generate the $(k,j)$th element of $\mathbf{A}^*$ using
%
\begin{eqnarray}
\label{eq:A1} a_{k,j}&=&u_{k,j}v_{k,j}, \qquad k=1,\dots,d, j=p+1,\dots,L,
\\
\label{eq:A} u_{k,j} &\sim&\mathcal{N}(0,1), \qquad v_{k,j} \sim\mathcal
{B}(1-\xi_j),
\end{eqnarray}
where $\mathcal{B}(\pi)$ is the Bernoulli distribution with
probability of $1$ equal to $\pi$. In the $(l+1)$th iteration, we let
%
\begin{equation}
\label{xi}
\xi^{(l+1)}_j \sim \operatorname{Beta}\biggl(\alpha,\frac{\alpha(1-\bar{\xi
}^{(l)})}{\bar{\xi}^{(l)}}\biggr), \qquad   j=p+1,\ldots,L,
\end{equation}
where $\xi^{(l+1)}_j$ is the sparsity of the $j$th column of $\mathbf
{A}^{*(l+1)}$. Note that $E(\xi^{(l+1)}_j)=\bar{\xi}^{(l)}$\vspace*{1pt} for all
values of $\alpha$. We found $\alpha=5$ produced a reasonable amount
of variance in sparsity levels.

We then combine these $L-p$ columns with the selected $p$ columns from iteration
$l$ to form the new intermediate transformation matrix $\mathbf{A}^{*(l+1)}$.
We select $\bar{\xi}^{(0)}=0.5$ for the initial sparsity level, which
seems to provide a reasonable compromise
between the variable selection and variable combination paradigms.

The full MASS algorithm is explicitly described as follows:
\begin{enumerate}
\item Set $\bar{\xi}^{(0)}=0.5$ and generate an initial $\mathbf
{A}^{*(0)}$ using (\ref{eq:A1}) through (\ref{xi}). Calculate
$\mathbf{Z}^{*(0)}$ by Equation (\ref{eq:linear}).
\item Select $p$ variables from $\mathbf{Z}^{*(0)}$ and keep the
corresponding $p$ columns from $\mathbf{A}^{*(0)}$ to obtain $\mathbf
{A}^{(0)}$.
\item Iterate until $l=I$.
\begin{enumerate}[(c)]
\item[(a)] Generate $L-p$ new directions $\mathbf{A}^*_{\mathrm{new}}$ by
using (\ref{eq:A1}) through (\ref{xi}).
\item[(b)] Let $\mathbf{A}^{*(l)}=(\mathbf{A}^{(l-1)},\mathbf
{A}^*_{\mathrm{new}})$ and use Equation (\ref{eq:linear}) to obtain $\mathbf
{Z}^{*(l)}$.
\item[(c)] Select $p$ variables from $\mathbf{Z}^{*(l)}$.
\item[(d)] Keep the corresponding $p$ columns from $\mathbf
{A}^{*(l)}$ to obtain $\mathbf{A}^{(l)}$.
\item[(e)] Calculate $\bar{\xi}^{(l)}$ for $\mathbf{A}^{(l)}$.
\item[(f)] Go to (a).
\end{enumerate}
\item Apply the selected classification technique to the final $\mathbf
{Z}^{(I)}$.
\end{enumerate}

\subsection{Theoretical justification}\label{sec2.3}

Here we show that MASS will asymptotically select the correct sub-space,
provided a ``reasonable'' variable selection method is utilized in Step
3(c). Assumption~\ref{var.select} below formally defines reasonable.
Suppose our variable selection method must choose among $\mathbf{z}_1,
\ldots, \mathbf{z}_L$
potential variables. Let $\mathbf{Z}_0 \in\mathbb{R}^{n\times p}$
represent the
$p$-dimensional set of true variables. Note $\mathbf{Z}_0$ is not
necessarily a
subset of $\mathbf{z}_1, \ldots, \mathbf{z}_L$. Define $\tilde
\mathbf{Z}_n\in\mathbb{R}^{n\times p}$ as
the $p$ variables among $\mathbf{z}_1,\mathbf{z}_2,\ldots,\mathbf
{z}_L$ that minimize $\|\tilde{\mathbf{Z}}_n - \mathbf{Z}_0\|^2$ for
a sample of
size $n$. Then we assume that the variable selection method chosen for
MASS satisfies the following
property:

\begin{assumption}
\label{var.select}
There exists some $\varepsilon>0$ such that, provided
%
\begin{equation}
\label{eq:assumption}
\frac{1}{n} \|\tilde\mathbf{Z}_n - \mathbf{Z}_0\|^2\le\varepsilon,
\end{equation}
then $\tilde\mathbf{Z}_n$ is chosen by the variable selection method
almost surely as
$n\rightarrow\infty$.
\end{assumption}

Assumption~\ref{var.select} is a natural extension of the definition
of consistency of a variable selection method, namely, that it
asymptotically selects the correct model. Here we extend this idea
slightly by assuming that, provided a set of candidate predictors that is
arbitrarily close to the true predictors is presented, the variable
selection method will asymptotically choose these variables.
Theorem~\ref{mass.theorem} provides some theoretical
justification for the MASS methodology.

\begin{theorem}
\label{mass.theorem}
Let $\tilde\mathbf{Z}_n^{(I)}$ represent the $p$ variables selected
by MASS after
performing $I$ iterations on a sample of size $n$. Then, under the
linear subspace model given by \textup{(\ref{eq:linear})} and \textup{(\ref{yeqn})},
provided a variable selection method is chosen such that
Assumption~\textup{\ref{var.select}} holds, as $n$ and $I$ approach infinity,
\[
\frac{1}{n} \big\|\tilde\mathbf{Z}_n^{(I)} - \mathbf{Z}_0\big\|
^2\rightarrow0\qquad   a.s.
\]
\end{theorem}

%

The proof of Theorem~\ref{mass.theorem} is given in Appendix~\hyperref[app:A]{A}. Theorem~\ref{mass.theorem} guarantees that, provided a
reasonable variable selection method is chosen, then the sub-space chosen
by MASS will converge to the ``true'' sub-space, in terms of mean squared
error, as $n$ and $I$ converge to infinity. Note,
Theorem~\ref{mass.theorem} does not assume that (\ref{eq:assumption})
holds; Only that, if it does hold, then asymptotically $\tilde\mathbf
{Z}_n$ will be
selected.

\subsection{The nonlinear generalization} \label{generalsec}

Recently, nonlinear reduction work has mostly concentrated on local
neighborhood methods such as Isomap \citet{TSL00} and LLE \citet{RS00}.
One limitation of these approaches is that they are clustering
based and hence are unsupervised methods. Another limitation
is that these approaches only consider local feature spaces. They perform
well when the data belong to a single well sampled cluster, but fail when
the points are spread among multiple clusters. MASS can also produce a
nonlinear reduction without the aforementioned problems.

Recall that MASS attempts to compute the $f_{j,k}$'s subject to
(\ref{smooth.constraint}) holding for some $\lambda\ge0$. It is
not hard to show that, among all functions that interpolate a given set of
points, the one that minimizes the integrated squared second derivative
will always be a natural cubic spline \citet{R67}. Hence, since we
wish to
choose a set of functions that reproduce the $z_{i,j}$'s subject to
(\ref{smooth.constraint}), it seems sensible to model each $f_{j,k}$ using
a $q$-dimensional natural cubic spline (NCS) basis, $\mathbf{b}(t)$.


Using this formulation, (\ref{eq:general}) becomes
$z_{i,j}=\sum_{k=1}^d
f_{j,k}(x_{i,k})=\sum^d_{k=1}\mathbf{b}(x_{i,k})^T\times \break \bolds{\theta
}_{j,k}$, where
$\bolds{\theta}_{j,k}$ represents the basis coefficients for
$f_{j,k}$. Since the
$z_{i,j}$'s are just linear functions of the $\bolds{\theta}$'s, we
can rewrite
(\ref{eq:general}) in the simpler linear form, (\ref{eq:linear}),
%
\begin{equation}
\label{gen.linear}
\mathbf{Z}=\mathbf{X}^*\Theta,
\end{equation}
where
$\mathbf{X}^*=(B(\mathbf{x}_1)|B(\mathbf{x}_2)|\cdots|B(\mathbf{x}_d))
\in\mathbb{R}^{n\times(q\times d)}$,
$\Theta=(\bolds{\theta}_1,\dots,\bolds{\theta}_p) \in\mathbb
{R}^{(q\times d)\times
p}$, $\bolds{\theta}_j=(\theta^{T}_{j,1},\dots,\theta
^{T}_{j,d})^T$, and
$B(\mathbf{x}_k)=(\mathbf{b}(x_{1,k}),\dots,\mathbf{b}(x_{n,k}))^T$.

The only complication in using the standard linear MASS methodology to fit
(\ref{gen.linear}) is ensuring that (\ref{smooth.constraint}) holds.
However, in Appendix~\ref{app:B} we show that some minor adaptations to
(\ref{eq:A}) ensure that (\ref{smooth.constraint}) holds for all candidate
$\theta$'s that we generate. This is one of the advantages of the
stochastic search process---it is easy to search only the feasible values
of the $\Theta$ space. In all other respects, the MASS methodology as
outlined in Section \ref{sec2.2} can be applied without any alterations to estimate
a non-linear sub-space of the data.
It should also be noted that Theorem~\ref{mass.theorem} can be
extended to
the nonlinear setting, provided we assume that the true nonlinear
sub-space satisfies (\ref{smooth.constraint}).



\subsection{Preliminary reduction}

MASS can, in general, be applied to any data. However, \citet{FL08}
argue that a successful strategy to deal with ultra-high dimensional data
is to apply a series of dimension reductions. In our setting this strategy
would involve an initial reduction of the dimension to a ``moderate''
level followed by applying MASS to the new lower dimensional data. This
two stage approach potentially has two major advantages. First,
\citet{FL08} show that prediction accuracy can be considerably improved by
removing dimensions that clearly appear to have no relationship to the
response. Our own simulations reinforce this notion. Second,
stochastic search algorithms such as MASS can require significant
computational expense. Reducing the data dimension before applying MASS
provides a large increase in efficiency.

In this paper we consider three methods for reducing the data into $m$
$(m>p)$ dimensions. The first approach is conventional PCA, which has the
effect of selecting the best $m$ dimensional space in terms of minimizing
the mean squared deviation with the original $d$ dimensional space. The
second approach is the sure independence screening (SIS) method based on
correlation learning \citet{FL08}. It computes the componentwise marginal
correlations between each predictor and the response vector. One then
selects the variables corresponding to the $m$ largest correlations. PCA
is an unsupervised variable combination approach, while SIS is a
supervised variable selection method. PCA may exclude important
information among variables with less variability, while SIS may tend to
select a redundant subset of predictors that have high correlations with
the response individually but are also highly correlated among themselves.
Our third dimension reduction approach, which we call PCA-SIS, attempts to
leverage the best of both PCA and SIS by first using PCA to obtain $n$
orthogonal components and then treating these components as the predictors
and using SIS to select the best $m$ in terms of correlation with the
response. PCA-SIS can be thought of as a supervised version of PCA. We
compare these three types of preliminary reduction methods, along with the
effect of performing no initial dimension reduction, in our simulation
studies.

\citet{FL08} argue that the dimension of the intermediate space $m$ should
be chosen as
%
\begin{equation} \label{eq:inter}
m=\frac{2n}{\log(n)}.
\end{equation}
As we have found that this approach generally works well, we have
adopted Equation (\ref{eq:inter}) for selecting $m$ in this paper.


\subsection{Implementation issues} \label{subsec:imp}

MASS requires the choice of a variable selection technique. In
principle, any
variable selection technique can be applied here. We considered several possible
methods. In the context of classification, a natural approach is to use a
GLM version of the Lasso using a logistic regression framework. We
examined this approach using the GLMpath methodology of \citet{PH07}.
Interestingly, we found that simply using the standard Lasso procedure to
select the variables gave similar levels of accuracy to GLMpath and
required considerably less computational effort. Therefore, in our
implementation of MASS we use the first $p$ variables selected by the Lars
algorithm \citet{EHJT04}. However, in practice, one could implement our
methodology with any standard variable selection method such as SCAD,
Dantzig selector, etc. There is an extensive literature examining the
circumstances under which the Lasso will asymptotically select the correct
model [see, e.g., \citet{FK00}, \citet{TG05}]. Hence, it seems reasonable to
suppose that
Assumption~\ref{var.select} in Section~\ref{sec2.3} will hold.

To implement MASS, one must choose values for the number of iterations,
$I$, the number of random columns to generate, $L$, and the final number
of columns chosen, $p$. In our experiments we used $I=500$ iterations. We
found this value guaranteed a good result and often fewer
iterations were required. In general, tracking the deviance of the
Lasso at
each iteration provided a reliable measure of convergence. The value of
$L$ influences the convergence speed of the algorithm as well as the
execution time. We found that the best results were obtained by choosing
$L$ to be a relatively large value for the early iterations but to
have it decline over iterations. In particular, we set $L=n/2$ for the
first iteration and
had it decline to $2p$ by the final iteration. This approach is similar
in spirit
to simulated annealing where the temperature is lowered over time. It
guarantees that the Lasso has more variables to choose from at the early
stages, but then as the search moves toward the optimum, decreasing $L$ will
improve the reliability of the selected $p$ variables in addition to
speeding up the algorithm.

The choice of $p$ can obviously have an important impact on the
results. In general, $p$ should be chosen as some balance of the
classification accuracy and the computational expense. One reasonable
approach is to use an objective criterion to choose $p$, such as
cross-validation (CV). In other situations, some prior knowledge can be applied
to select $p$. For example, in the fMRI study that we examine in
Section~\ref{sec:3}, prior knowledge and assumptions on the regions of
interest can be used to decide on a reasonable value for $p$.

\section{Applications} \label{sec:3}
In this section we apply linear MASS to two data sets from real
studies: an fMRI data set and a gene microarray data set. As a
comparison to MASS, we also apply classic logistic regression (LR) or
support vector machine (SVM) to the lower dimensional data produced
using a straight Lasso method (Lars), a generalized Lasso method
(GLMpath), SIS and PCA. They all utilize equation (\ref{eq:linear})
but compute ${\bf A}$ directly using their own methodologies. In both
studies, we use 500 iterations for MASS.

\subsection{fMRI brain imaging data} \label{subsec:fmri}
The fMRI data was obtained from the imaging center at the University of
Southern California. The raw data consisted of 200 3-D brain images
recording the blood oxygen level dependent (BOLD) response for a
subject who was conducting a visual task. After preprocessing, each
image contained 11,838 voxels. One research question was to divide the
200 images into task (96) and baseline (104) images based on the 11,838
voxels. To answer this question, we randomly divided the data to
training (150) and test (50) samples. Since $d \gg n$, a preliminary
reduction was needed. The intermediate scale was \mbox{$m=60$} by Equation
(\ref{eq:inter}). We tested $p=10,20,30,40,50,60$, which were
considered to be good balances of the execution time and the
classification accuracy. SVM was used as the base classifier.

\begin{table}
\caption{Test MCR$^*$ and $p^*$ on fMRI data using SVM}
\label{tbl:fmriSVM}
\begin{tabular*}{\textwidth}{@{\extracolsep{\fill}}lcc@{}}
\hline
\textbf{Methods} & \textbf{Test MCR$\bolds{^*}$} & \textbf{Optimal $\bolds{p^*}$} \\
\hline
Lars & 0.20 & 50 \\
GLMpath & 0.22 & 50 \\
SIS & 0.40 & 30 \\
PCA & 0.26 & 40 \\
SIS-Lars & 0.36 & 30 \\
PCA-Lars & 0.22 & 40 \\
PCA-SIS-Lars & 0.12 & 50 \\
SIS-GlMpath & 0.38 & 30 \\
PCA-GLMpath & 0.22 & 40 \\
PCA-SIS-GLMpath & 0.10 & 50 \\
SIS-MASS & 0.373 (0.008) & 50 \\
PCA-MASS & 0.155 (0.005) & 40 \\
PCA-SIS-MASS & 0.042 (0.005) & 50 \\
\hline
\end{tabular*}
\end{table}

\begin{figure}[b]

\includegraphics{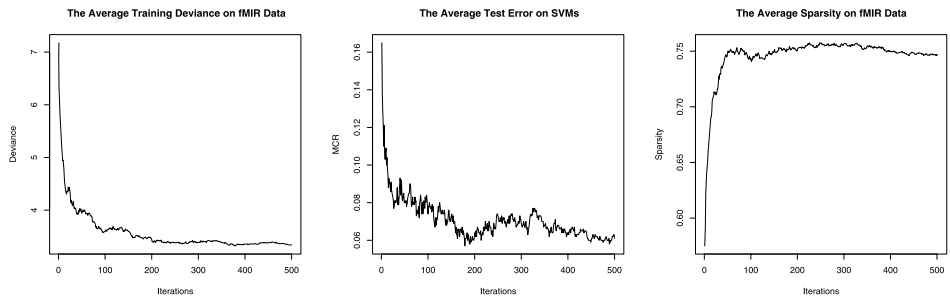}

\caption{Preliminary reduction by PCA-SIS-MASS on fMRI data $(p=30)$.}
\label{fig:fmri}
\end{figure}

Table \ref{tbl:fmriSVM} reports the minimum test MCR, MCR$^*$, and its
corresponding $p$, $p^*$, for each method. For each $p$, we applied
MASS twenty times on the training data and obtained the average MCR on
the test data. We then reported the minimum average test MCR, and its
corresponding $p$. In this table as well as in the rest of the paper,
SIS-MASS means that we first used SIS to reduce the data and then
applied MASS. Similarly, PCA-SIS-Lars means that PCA-SIS was the first
reduction method and then Lars was applied, etc.

Obviously, PCA-SIS-MASS dominates other methods. We show the training
deviance, the average test MCR and the average sparsity as some
function of the number of iterations in Figure \ref{fig:fmri}. As we
can see, the training deviance and the average test MCR decline rapidly
and then level off, indicating the model\vspace*{1pt} improves quickly. The average
$\bar{\xi}$ path for PCA-SIS-MASS indicates MASS chose a relatively
sparse matrix as the optimal transformation matrix. We also examined
the relative performance of MASS in comparison to a representative
sampling of some modern classification methods [Support Vector Machines
(SVM), \mbox{$k$-Nearest} Neighbors (kNN), Neural Networks (NN) and Random
Forests (RF)] on the \mbox{$m$-dimensional} preliminary reduced data. As shown
in Table \ref{tbl:fmriclass}, all four methods were inferior to MASS.


\subsection{Leukemia cancer gene microarray data}
\begin{figure}

\includegraphics{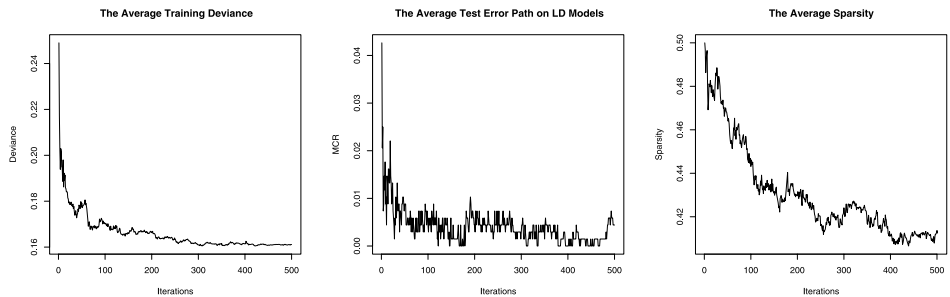}

\caption{PCA-SIS-MASS on leukemia cancer data.}
\label{fig:leuk}
\end{figure}

The second real-world data set was the leukemia cancer data from a gene
microarray study by \citet{G99}. This data set contained $72$ tissue
samples, each with 7129 gene expression measurements and a cancer type
label. Among the $72$ tissues, $25$ were samples of acute myeloid
leukemia (AML) and $47$ were samples of acute lymphoblast leukemia
(ALL). Within the $38$ training samples, $27$ were ALL and $11$ AML.
Within the $34$ test samples, $20$ were ALL and $14$ AML. In some
previous studies \citet{FF08}, \citet{FL08}, $16$ genes were ultimately chosen. In
this study, we also picked $p=16$ genes for MASS. However, for the
other counterpart methods, we examined all possible $p$'s, that is,
$p=1,2,\ldots,21$, to obtain MCR$^*$ and their corresponding $p^*$'s.
This is considered to be an advantage for the counterpart methods and a
disadvantage for MASS. The intermediate dimension was chosen to be
$m=21$ by Equation (\ref{eq:inter}). A LR model was applied for
classification. Similar to the fMRI study, PCA-SIS-MASS provided the
lowest MCR (see Table \ref{tbl:leuk}), and it was better than the
nearest shrunken centroids method mentioned in \citet{THNC02}, which
obtained a MCR of $2/34=0.059$ based on $21$ selected genes \citet
{FL08}. Figure \ref{fig:leuk} shows the resulting graphs. The sparsity
of the transformation matrix leveled off at about $0.41$.

\begin{table}
\caption{Test MCR on the preliminary reduced data by different classifiers}
\label{tbl:fmriclass}
\begin{tabular*}{\textwidth}{@{\extracolsep{\fill}}lcccc@{}}
\hline
& \textbf{SVM} & \textbf{kNN} & \textbf{NN} & \textbf{RF} \\
\hline
SIS- & 0.36 & 0.40 & 0.44 & 0.34 \\
PCA- & 0.26 & 0.26 & 0.38 & 0.22 \\
PCA-SIS- & 0.24 & 0.20 & 0.36 & 0.20 \\
\hline
\end{tabular*}
\end{table}

\begin{table}[b]
\tabcolsep=0pt
\caption{Test MCR$^*$ and corresponding $p^*$ or standard errors (in
the parentheses) on leukemia cancer~data~using LR}
\label{tbl:leuk}
\begin{tabular*}{\textwidth}{@{\extracolsep{\fill}}ld{2.24}d{1.24}c@{}}
\hline
&  \multicolumn{1}{c}{\textbf{Lars}} & \multicolumn{1}{c}{\textbf{GLMpath}} & \textbf{MASS ($\bolds{p=16}$)}\\
\hline
$m=21$ & &&  \\
 \quad SIS-& 0.147\ (p^*=12,14\mbox{--}17,19,20)&0.088\ (p^*=16,17,20,22\mbox{--}24)& 0.176 (0.010) \\
\quad  PCA- & 0.029\ (p^*=13,15\mbox{--}17) & 0.029\ (p^*=9) & 0.056 (0.008) \\
\quad  PCA-SIS- & 0.029\ (p^*=16) & 0.029\ (p^*=15) & 0.004 (0.002) \\
\hline
\end{tabular*}
\end{table}

\section{Simulation studies} \label{sec:4}

In this section we further investigate the performance of MASS under
different conditions. We give five simulation examples. The first
simulation investigates the nonlinear setting, while the remaining simulations
concentrate on the linear case. In all simulations we show results with
MASS applied using the LR classifier and the SVM classifier. In
addition, we also apply LR and SVM to the original full dimensional
data (FD) and to lower
dimensional data produced using Lars, GLMpath, SIS and PCA.

\subsection{Simulation study I: Nonlinear data}

In this setting we simulated data using the nonlinear additive model
assumed by MASS. We first selected an NCS basis and then generated
$\theta_{j,k}$'s such that (\ref{smooth.constraint}) holds for a given
$\lambda$. We used the same generation scheme as that used by the MASS
algorithm. Details of the generation process are provided in
Appendix~\ref{app:B}.

We simulated 20 training samples, each containing $100$ observations and
$50$~predictor variables generated from $\mathcal{N}(0,\mathbf{I})$.
The binary response variable was associated with the true $\mathbf
{Z}_0 \in\mathbb{R}^{n \times p_0}$ using a standard logistic
regression model. Here we let $p_0=2$. For each training data set a
corresponding set of $1000$
observations was generated as a test sample. In order to demonstrate the
importance of the sparsity of the transformation matrix to the
classification performance, we also implemented a modified version of MASS
where we fixed the sparsity of the transformation matrix, $\bar{\xi
}$, so
that it did not change over iterations. We call this method the
multivariate fixed stochastic search (MFSS) method. Presumably, MFSS with
a good value of sparsity will perform better than MASS, because MASS has
to adaptively search for $\bar{\xi}$. For MASS and MFSS, we let
$p=p_0=2$ dimensions.

We considered two scenarios, where the true curvatures were $\lambda
_0=5$ and $\lambda_0=10$. According to Equation (\ref
{smooth.constraint}), $\lambda$ is one of the indicators of the model
complexity. With a larger $\lambda$, the model is more complex. The
true sparsity for both scenarios were $\bar{\xi}_0=0.3$. The
counterpart methods are Lars and PCA with their optimal $p^*$'s. We
examined all possible values for $p$, that is, $p=1,2,\ldots,50$, for
Lars and PCA and report their MCR$^*$ in each simulation run. Table
\ref{tbl:nonlinear} shows the average test MCR using LR and SVM
classifiers. When we let $\lambda=\lambda_0$, MASS and MFSS
constantly produced good results. In particular, MFSS with $\lambda_0$
was significantly better than other methods. MFSS with incorrect
$\lambda$'s were inferior to MFSS with the correct $\lambda$, but
were still superior to Lars and PCA. Lars and PCA each suffered from an
inability to match the nonlinear structure of the data even with the
optimal $p^*$. When the model is very complex (e.g., $\lambda=10$),
overfitting may occur in MASS even with a small number of iterations.
We discuss the potential problem of overfitting in Section~\ref{sec:5}.

\begin{table}
\caption{Average test MCR (standard errors) in Simulation I}
\label{tbl:nonlinear}
\begin{tabular*}{\textwidth}{@{\extracolsep{\fill}}lcccc@{}}
\hline
&\multicolumn{2}{c}{$\bolds{\lambda_0=5,\bar{\xi}_0=0.3}$}&\multicolumn{2}{c@{}}{$\bolds{\lambda_0=10,\bar{\xi}_0=0.3}$} \\[-6pt]
&\multicolumn{2}{c}{\hrulefill}&\multicolumn{2}{c@{}}{\hrulefill} \\
& \textbf{LR} & \textbf{SVM} & \textbf{LR} & \textbf{SVM} \\
\hline
FD & 0.424 (0.005) &0.414 (0.005)&0.412 (0.006)& 0.389 (0.007) \\
Lars (with $p^*$) & 0.398 (0.006) &0.391 (0.007) &0.407 (0.007)&0.405 (0.008) \\
PCA (with $p^*$) & 0.424 (0.008) &0.426 (0.008) &0.487 (0.005)&0.489 (0.004) \\
MASS $(\lambda=\lambda_0)$ & 0.234 (0.008) &0.239 (0.007)&0.302 (0.007)&0.300 (0.006) \\
MFSS $(\lambda=0, \bar{\xi}=0.3)$ & 0.289 (0.009) &0.284 (0.008)&0.352 (0.006) &0.351 (0.007) \\
MFSS $(\lambda=5, \bar{\xi}=0.3)$ & 0.222 (0.008) &0.230 (0.008)&0.315 (0.005) &0.309 (0.006)\\
MFSS $(\lambda=10, \bar{\xi}=0.3)$ & 0.378 (0.007) &0.392 (0.008)&0.279 (0.006) &0.275 (0.005)\\
\hline
\end{tabular*}
\end{table}

\subsection{Simulation study II: Sparse $\mathbf{A}$ case}

This simulation was designed to examine the performance of MASS in a
sparse model situation where the response was only associated with a
handful of predictors. The training and test samples were generated
from $\mathcal{N}(\mathbf{0},\Sigma)$, where the diagonal elements
of $\Sigma$ were 1 and the off-diagonal elements were 0.5. We then
rescaled the first $p_0=5$ columns to have a standard deviation of
$10$. We call these columns with the most variability the \textit{major
columns}, and the rest the \textit{minor columns}.

We tested two scenarios by creating two different true transformation
matrices. In the first scenario the true transformation matrix was
$\mathbf{A}_0=\operatorname{perm}_1([\mathbf{I}_{p_0},\mathbf{0}]^T)$,
where $\operatorname{perm}_1$ was the row permutation that made the unit row
vectors in $\mathbf{A}_0$ associate with the major columns of the data
matrix. In the second scenario the true transformation matrix was
$\mathbf{A}_0=\operatorname{perm}_2([\mathbf{I}_{p_0},\mathbf{0}]^T)$,
where $\operatorname{perm}_2$ was the row permutation that made the unit row
vectors in $\mathbf{A}_0$ associate with any $p_0$ of the minor
columns. In both scenarios, the true dimension of the subspace is
$p_0=5$. Again, the group labels were generated by a standard logistic
regression model:
%
\begin{equation} \label{eq:LR}
\Pr(y_i=1|Z_i)=\frac{e^{Z_i^T\beta}}{1+e^{Z_i^T\beta}},
\end{equation}
where $Z_i$ is the $i$th point in the reduced space and $\beta$ is a
$p_0$-dimensional coefficients vector of the logistic regression. The
elements of $\beta$ are generated from some uniform distributions,
that is, $\mathcal{U}(-0.5,0.5)$ for scenario 1 and $\mathcal
{U}(-2,2)$ for scenario 2, in order to make the Bayes error rates for
both scenarios remain roughly around 0.1. We expect that PCA should
perform best in scenario 1 because this case matches the PCA working
mechanism exactly; on the other hand, it should perform poorly in
scenario 2 because the first $p$ eigenvectors tended to concentrate on
the major columns where no group information resides.

\begin{figure}

\includegraphics{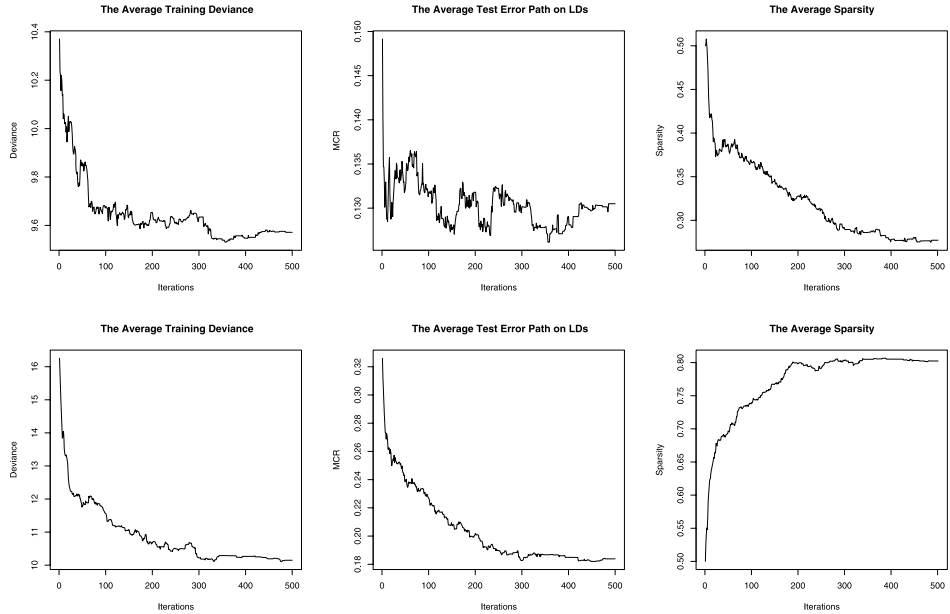}

\caption{MASS performance graphs in Simulation II. Upper panels are
scenario 1 and lower panels scenario 2.}
\label{fig:case1}
\end{figure}

We ran Lars, GLMpath, SIS and PCA for multiple values of $p$, and for
Lars, GLMpath and SIS, $p=5$ is the best value, $p^*$. So we reported
the average test error when $p=5$ for these methods. We mandatorily
assigned $p=5$ to MASS and MFSS (with $\bar{\xi}=\bar{\xi
}_0=0.98$). This was a disadvantage for these methods. We also
calculated the average Bayes rates. As is shown in Table \ref
{tbl:sim}, not surprisingly, LR generally outperformed SVM on this data
because the data were generated using a logistic regression model. As
expected, PCA worked well in scenario 1 but poorly in scenario 2. One
of the reasons the performance of PCA is so poor in scenario 2 is that
we used $p=5$, not the $p^*$. If the $p^*$ is used, the performance of
PCA in scenario 2 may be improved. The supervised methods had an
advantage in the latter scenario because they always looked for
$\mathbf{A}$'s with high predictive power. MASS and MFSS performed
well in both scenarios. Note that since the true model was highly
sparse, the variable selection methods (Lars and GLMpath) performed
well too. As we can see from Figure \ref{fig:case1}, the average test
MCR and deviance in both scenarios were decreasing, which means MASS
tended to choose ``good'' directions from $\mathbf{A}^*$ over
iterations, and hence, the classification accuracy was improved gradually.

It is interesting to see that the sparsity levels for the estimated
projection matrices of these two scenarios were different, although the
sparsity for the true projection matrices were in fact the same: $\bar
{\xi}_0=0.98$. One possible reason for the difference is that while
$\mathbf{A}$'s may have elements close to zero which make little
contribution in extracting variables, these elements still contribute
to its ``denseness'' according to our sparsity definition. Another
possible reason is that for scenario 1, where all information was
located in major columns, the noise level is low. The inclusion of
minor columns does not highly influence classification accuracy.
However, in scenario 2, since the noise level is high, the search must
find the correct sparsity in order to improve accuracy. This explains
why MFSS with $\bar{\xi}=0.98$ performed worse than MASS in scenario
1 but better in scenario 2.


\begin{table}
\tabcolsep=0pt
\caption{Bayes rates and average test MCR in Simulation II}
\label{tbl:sim}
\begin{tabular*}{\textwidth}{@{\extracolsep{\fill}}lccccccccc@{}}
\hline
\textbf{Scenario} & \textbf{Classifier} & \textbf{Bayes} & \textbf{FD} & \textbf{Lars} & \textbf{GLMpath} & \textbf{SIS} & \textbf{PCA} & \textbf{MASS}& \textbf{MFSS} \\
\hline
1&LR&0.114& 0.268&0.150 &0.149 &0.192 &0.114 &0.130 & 0.154\\
& &(0.007)&(0.011)&(0.012)&(0.012)&(0.021)&(0.009)&(0.007)&(0.008)\\
[3pt]
&SVM& & 0.254 &0.166 &0.166 &0.213 &0.139 &0.136 &0.157 \\
& &&(0.017)&(0.013)&(0.013)&(0.022)&(0.008)&(0.006) &(0.008) \\
[6pt]
2&LR&0.112& 0.273&0.143 &0.148 &0.217 &0.477 &0.184 &0.141\\
& &(0.006)&(0.009)&(0.009)&(0.010)&(0.017)&(0.006)&(0.012)&(0.006)\\
[3pt]
&SVM& &0.255 &0.164 &0.169 &0.242 &0.485 &0.189&0.152 \\
& &&(0.018)&(0.010)&(0.011)&(0.019)&(0.007)&(0.012) &(0.006) \\
\hline
\end{tabular*}
\end{table}

\subsection{Simulation study III: Dense $\mathbf{A}$ case}

In this simulation we investigated a dense model. We simulated the
input data $\mathbf{X}$ (the same size as in Simulation~II) from
$\mathcal{N}(0,\Sigma)$, where the diagonal elements of $\Sigma$
were 1 and the off-diagonal elements were 0.5. Unlike the sparse case,
the elements of the true 5-dimensional $\mathbf{A}_0$ were generated
from the standard normal distribution. Since the $\mathbf{A}_0$ matrix
is dense, the columns in $\mathbf{Z}$ will be linear combinations of
columns in $\mathbf{X}$. If we use a linear LR model (\ref{eq:LR}) to
generate group labels as in Simulation II, the generated group labels
will depend on a linear combination of the columns of $\mathbf{Z}$.
Therefore, the group labels are actually directly related to a linear
combination of all the columns of $\mathbf{X}$, thus, removing the
concept of a lower dimensional space. Hence, to ensure the response was
a function of the lower dimensional space, we generated the group
labels using a nonlinear logistic regression,
%
\begin{equation} \label{eq:nonlr}
\Pr(y_i=1|Z_i)=\frac{e^{g(Z_i)}}{1+e^{g(Z_i)}},
\end{equation}
where $g(Z_i)=\sin(0.05\pi Z_i)^T\bolds{\beta}$. This guarantees
that most $0.05\pi Z_i$ values fall into the range of $(-\pi/2, \pi
/2)$. Hence, the nonlinearity is produced by that particular part of a
sine function.

\begin{figure}[b]

\includegraphics{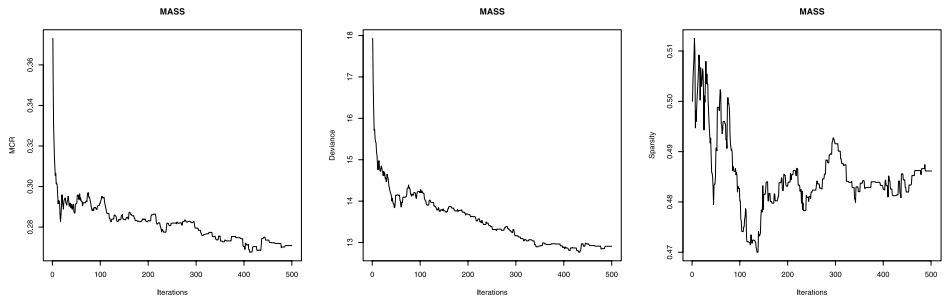}

\caption{MASS performance graphs in Simulation III with $p=5$.}
\label{fig:mass3}
\end{figure}

\begin{figure}

\includegraphics{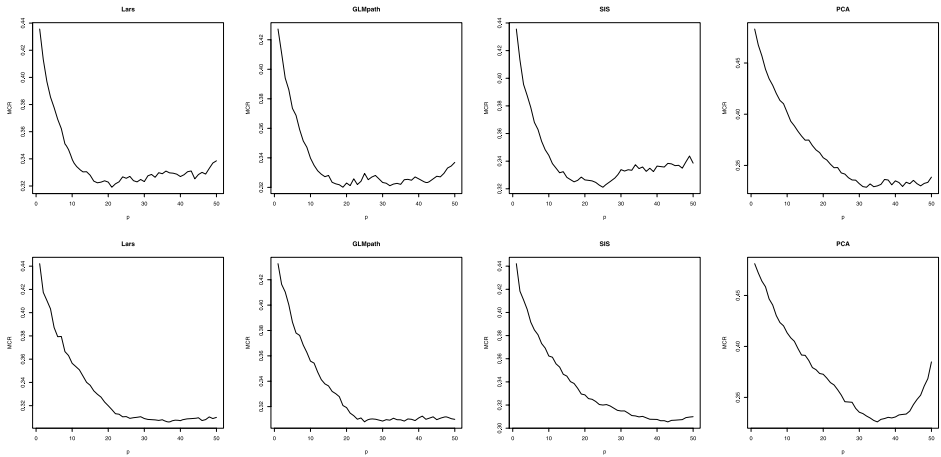}

\caption{Average test MCR for all values of $p$ in Simulation III.
Upper panels: LR; lower panels: SVM.}
\label{fig:CV3}
\end{figure}

Since the data scale was moderate and all the variables contain group
information, the optimal $p^*$ for different methods may be different
and not always be $p_0$. We reported the average minimal MCR$^*$'s and
its corresponding $p^*$'s for Lars, GLMpath, SIS and PCA, respectively.
As to MASS and MFSS, we still fix $p=5$, which is considered a
disadvantage for these two but an advantage for other methods. Since
$\mathbf{A}_0$ is dense, we assigned $\bar{\xi}=0$ to MFSS. Table
\ref{tbl:sim3} lists the average MCR$^*$ and $p^*$'s based on 20 pairs
of training and test data. Figure \ref{fig:mass3} shows the MASS
performance graphs. As can be seen, MASS and MFSS are still superior
when all the methods use their optimal $p^*$'s, with MFSS providing the
best results. Figure \ref{fig:CV3} shows the test MCR values by Lars,
GLMpath, SIS and PCA with different $p$'s. These methods all tend to
use larger numbers of variables.
%

\begin{table}
\caption{Average minimum MCR from Simulation III}
\label{tbl:sim3}
\begin{tabular*}{\textwidth}{@{\extracolsep{\fill}}lcccc@{}}
\hline
& \multicolumn{2}{c}{\textbf{LR}} & \multicolumn{2}{c@{}}{\textbf{SVM}} \\[-6pt]
& \multicolumn{2}{c}{\hrulefill} & \multicolumn{2}{c@{}}{\hrulefill} \\
& $\bolds{p^*}$ & \textbf{MCR}$\bolds{^*}$ & $\bolds{p^*}$ & \textbf{MCR}$\bolds{^*}$ \\
\hline
Bayes& \multicolumn{4}{c@{}}{0.082 (0.002)} \\
FD & --- & 0.339 (0.006) & --- & 0.317 (0.007) \\
Lars& 21 & 0.319 (0.006) & 37& 0.306 (0.006) \\
GLMpath & 19 & 0.320 (0.006) & 25& 0.308 (0.005) \\
SIS & 25 & 0.321 (0.005) & 43& 0.305 (0.006) \\
PCA & 32 & 0.329 (0.005) & 35& 0.326 (0,006) \\
[6pt]
MASS($p=5$)& &0.271 (0.010) & & 0.245 (0.008) \\
MFSS($p=5$)& &0.239 (0.008) & & 0.212 (0.007) \\
\hline
\end{tabular*}
\end{table}

\subsection{Simulation study IV: Contaminated data}

In order to examine the robustness of MASS against outliers, we created a
situation where the distributional assumptions were violated. We
generated the input data from a multivariate $g$-and-$h$ distribution
\citet{FG06} with $\mathbf{g}=\mathbf{h}=(0.5,\dots,0.5)^T \in\mathbb
{R}^{50}$. As with Simulation III, we used Equation (\ref{eq:nonlr})
to create the group labels, except that here we used $g(Z_i)=\sin
(0.005\pi Z_i)^T\bolds{\beta}$. Since this is a highly skewed and
heavy-tailed case, there are many extreme values. Hence, the data
domain is rather large. We use $\sin(0.005\pi Z_i)$ so that most
$0.005\pi Z_i$ values fall into the range of $(-\pi/2, \pi/2)$. As
with Simulation III, only half period of a sine curve is effective.

MASS and MFSS were performed with a fixed $p=5$, while all other
methods used their $p^*$'s. MFSS was implemented with $\bar{\xi}=0$.
The results are listed in Table~\ref{tbl:sim4}. As with Simulation
III, MASS and MFSS with a fixed $p=5$ outperformed other methods with
their $p^*$'s. All other methods performed poorly on these data.
Compared to Simulation III, the performance of MASS and MFSS did not
decline as much as other methods. This demonstrates that the proposed
method is not very sensitive to outliers.

\begin{table}[b]
\caption{Average minimum MCR from Simulation IV}
\label{tbl:sim4}
\begin{tabular*}{\textwidth}{@{\extracolsep{\fill}}lcccc@{}}
\hline
& \multicolumn{2}{c}{\textbf{LR}} & \multicolumn{2}{c@{}}{\textbf{SVM}} \\[-6pt]
& \multicolumn{2}{c}{\hrulefill} & \multicolumn{2}{c@{}}{\hrulefill} \\
& $\bolds{p^*}$ & \textbf{MCR}$\bolds{^*}$ & $\bolds{p^*}$ & \textbf{MCR}$\bolds{^*}$ \\
\hline
Bayes& \multicolumn{4}{c@{}}{0.068 (0.002)} \\
FD & --- & 0.482 (0.006) & --- & 0.490 (0.007) \\
Lars& 29 & 0.403 (0.004) & 20& 0.391 (0.004) \\
GLMpath & 37 & 0.401 (0.005) & 48& 0.389 (0.006) \\
SIS & 39 & 0.410 (0.005) & 31& 0.392 (0.005) \\
PCA & 35 & 0.412 (0.005) & 31& 0.399 (0.005) \\
[6pt]
MASS($p=5$)& &0.294 (0.006) & & 0.285 (0.007) \\
MFSS($p=5$)& &0.273 (0.005) & & 0.266 (0.007) \\
\hline
\end{tabular*}
\end{table}

\subsection{Simulation study V: Ultra high dimensional data}
In this section we simulated an ultra high dimensional situation
where $n=100,~d=1000$. Thus, a preliminary dimension reduction was needed.
Among the $1000$ variables, only 50~contained the group information.
These 50 informative variables were generated from a multivariate
normal distribution with variance 1 and correlation 0.5. The~950~noise
variables were generated from $\mathcal{N}(0,0.5\mathbf{I})$ in order
to achieve a reasonable signal-to-noise ratio. Letting $p_0=5$, the
$\mathbf{Z}_0$ was generated from the 50 informative variables through
a dense $\mathbf{A}_0$, with each element generated from the standard
normal distribution. The~group labels were still generated by Equation~(\ref{eq:nonlr}).

\begin{figure}[b]

\includegraphics{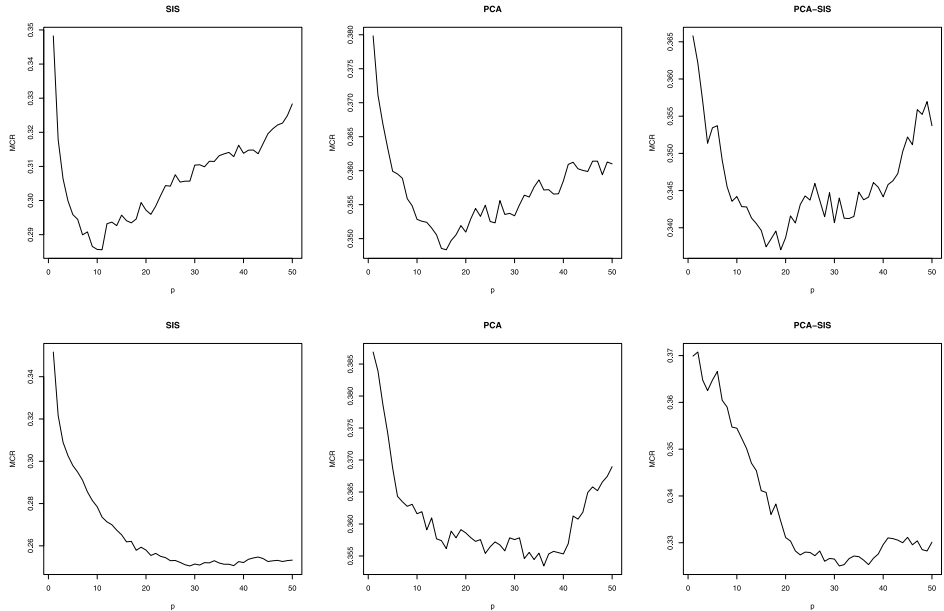}

\caption{Average test MCR for all values of $p$ on the first reduced
data by Lars in Simulation V. Upper panels: using LR; lower panels:
using SVM.}
\label{fig:CVhigh}
\end{figure}

We used the three aforementioned preliminary reduction methods, SIS,
PCA and PCA-SIS, to reduce the data into $m=50$ dimensions. Presumably,
if the preliminary reduction can extract the $m$ informative variables,
MASS and MFSS (with $\bar{\xi}=0$) will perform well. We also
implemented Lars and GLMpath on the preliminary reduced data. For Lars
and GLMpath, we reported their average test MCR$^*$ and $p^*$'s, while
for MASS and MFSS we still fixed $p=5$. Figure \ref{fig:CVhigh} shows
the average test MCR from different values of $p$ on the preliminarily
reduced data using Lars as the further reduction method. The test MCR
for GLMpath looks similar to Lars and thus is not shown. As we can see,
no matter what preliminary reduction method is applied, Lars tends to
choose larger values of $p^*$ than $p_0$. However, even with the
$p^*$'s, the test MCR$^*$ are above 0.250. In particular, when PCA and
PCA-SIS are applied, MCR$^*$ are above 0.320. SIS seems to be a better
preliminary reduction method for Lars and GLMpath in this simulation.

The average test MCR, using LR and SVM classifiers, are shown in
Table~\ref{tbl:high}. As we can see, all six implementations of MASS
and MFSS were statistically significantly better than Lars and GLMpath
with their $p^*$'s. In particular, SIS-MASS and PCA-SIS-MASS generally
provided the best results. It is not surprising that PCA or PCA-SIS as
the preliminary reduction methods did not fail, because the
signal-to-noise ratio was not large enough in this study. This
indicates that the improvement of the classification accuracy by MASS
is not only caused by the preliminary reduction but more by the method itself.

In order to further demonstrate that the improvement of the
classification accuracy was not due mainly to the preliminary reduction
but to the MASS method, we also applied SVM, kNN, NN and RF on the
50-dimensional reduced data without performing any further reduction.
The results are shown in Table \ref{tbl:class}. As can be seen, MASS
and MFSS performed extremely well relative to those approaches.

\begin{table}
\tabcolsep=0pt
\caption{Average test MCR (standard errors) in Simulation V}
\label{tbl:high}
\begin{tabular*}{\textwidth}{@{\extracolsep{\fill}}lccccc@{}}
\hline
&\textbf{Classifier} & \textbf{Lars (with $\bolds{p^*}$)} & \textbf{GLMpath (with $\bolds{p^*}$)} & \textbf{MASS} & \textbf{MFSS} \\
\hline
SIS- &LR &0.286 (0.012)&0.288 (0.012)&0.124 (0.007)&0.150 (0.009) \\
&SVM&0.250 (0.015) &0.253 (0.015)&0.121 (0.008)&0.155 (0.008) \\
[3pt]
PCA- &LR &0.348 (0.015)&0.347 (0.013)&0.165 (0.011)&0.182 (0.011) \\
&SVM&0.353 (0.014) &0.356 (0.013)&0.157 (0.011)&0.179 (0.012) \\
[3pt]
PCA-SIS-&LR &0.337 (0.013)&0.335 (0.012)&0.119 (0.010)&0.125 (0.011) \\
&SVM&0.325 (0.011) &0.327 (0.010)&0.102 (0.011)&0.119 (0.010) \\
\hline
\end{tabular*}
\end{table}


\begin{table}[b]
\caption{Average test MCR on the 50-dimensional first reduced data
using different classifiers}
\label{tbl:class}
\begin{tabular*}{\textwidth}{@{\extracolsep{\fill}}lcccc@{}}
\hline
& \textbf{SVM} & \textbf{kNN} & \textbf{NN} & \textbf{RF} \\
\hline
SIS- & 0.349 (0.017) & 0.356 (0.015) & 0.365 (0.016) & 0.344 (0.017) \\
PCA- & 0.358 (0.013) & 0.377 (0.013) & 0.323 (0.013) & 0.350 (0.015) \\
PCA-SIS- & 0.359 (0.012) & 0.364 (0.014) & 0.318 (0.012) & 0.351 (0.015) \\
\hline
\end{tabular*}
\end{table}

\section{Computational issues} \label{sec:5}

In this section we discuss several issues associated with MASS,
including the solution variability, the computational issue and the
potential problem of overfitting.

Regarding the randomized nature of MASS, one issue is the variability
of the solution, that is, the variability of the selected $\mathbf{A}$
for a fixed training and test pair from different simulation runs.
Theoretically, a space is determined by its orthogonal basis. We
presume two spaces are the same as long as they have the same basis
even though they can be rotated differently. Therefore, $\mathbf{A}$'s
can be treated the same if the $\mathbf{Z}$'s they produce have the
same principal components (PC). Hence, we examine the first PC of
$\mathbf{Z}$'s produced by different simulation runs.

We examine three different cases. The first two are the two scenarios
in Simulation II and the third is the fMRI data in Section~\ref
{subsec:fmri}. For each of the first two cases, we generate a training
data set of 100 observations paired with a test data set of 1000
observations. In each paired data set, we apply MASS 100 times and
obtain 100 $\mathbf{Z}$'s. We then extract the first PC for each
$\mathbf{Z}$ on the test data and calculate the average absolute
pairwise correlations between MASS runs:
\[
\overline{|\rho_{\mathrm{PC1}}|}=\frac{1}{4950}\sum_{s<t}|\rho_{s,t}|,
\]
where $\rho_{s,t}$ is the correlation of the first PC between the
$s$th and the $t$th runs.

For the fMRI data, we fixed the training data (150 observations) and
test data (50 observations), and conducted a preliminary reduction
using PCA-SIS to reduce the data space to 60 dimensional. The
$\overline{|\rho_{\mathrm{PC1}}|}$ was also calculated. In the first two
cases, we set $p=5$ and in the third, we set $p=50$.

Table \ref{tbl:A} shows $\overline{|\rho_{\mathrm{PC1}}|}$, the proportion of
variance that the first PC contains, and the standard errors of test
MCR from 100 runs. In all cases, $\overline{|\rho_{\mathrm{PC1}}|}$'s are
reasonably high (all above 0.70). Particularly, in the first case,
$\overline{|\rho_{\mathrm{PC1}}|}=0.968$, which indicates the first PC
accounting for $89\%$ of the data variance, are highly consistent based
on 100 simulation runs. PCs containing larger variance are more
consistent than PCs with less variance. In addition, the SE for MCR are
fairly small for all the cases (e.g., 0.006, 0.007 and 0.005,
respectively). All these indicate the solution produced by MASS is
fairly stable.

\begin{table}[b]
\caption{Check of solution stability: $\overline{|\rho_{\mathrm{PC1}}|}$'s
(standard errors), proportion of variance (standard errors) and~standard errors for MCR}
\label{tbl:A}
\begin{tabular*}{\textwidth}{@{\extracolsep{\fill}}lccc@{}}
\hline
& \textbf{PC1} & \textbf{Variance ($\bolds{\%}$)} & \textbf{SE for MCR} \\
\hline
Simulation II (1) & 0.968 (0.000) & 89\% (0.004) & 0.006 \\
Simulation II (2) & 0.830 (0.001) & 74\% (0.012) & 0.007 \\
fMRI & 0.732 (0.002) & 42\% (0.013) & 0.005 \\
\hline
\end{tabular*}
\end{table}

We compared MASS on the fMRI data, in terms of both classification
accuracy and the computational time, to the SA-Sparse method by \citet
{TWJ09}, which is also a stochastic search procedure. The approach of
\citet{TWJ09} provides a useful comparison because it also uses a
stochastic search but implements the search process using a simulated
annealing approach. The code was written in R and run on a Dell Precision
workstation (CPU 3.00~GHz; RAM 2.99~GHz, 16.0~GB). We recorded the total
CPU time (in seconds) of the R program for both methods. PCA was
applied first to reduce the data dimension to $m=80$ and then both
methods searched for a $p=30$ dimensional subspace. Table~\ref
{tbl:time} reports the time consumed and the classification accuracy on
the test data when both methods use 500~iterations. Both MASS and
SA-Sparse took a similar time period to run.
The key difference was that MASS converged well before $500$~iterations,
while SA-Sparse did not. Hence, the error rate for SA-Sparse was much
higher. SA-Sparse took
approximately $5000$~iterations to converge, which resulted in an order of
magnitude more computational effort. In addition, even when $5000$
iterations were chosen for SA-Sparse, the average test errors were
still higher
than for the MASS method.

\begin{table}
\caption{Execution time for each run and the average test MCR on fMRI
data using SVM}
\label{tbl:time}
\begin{tabular*}{\textwidth}{@{\extracolsep{\fill}}lcc@{}}
\hline
& \textbf{CPU Time/Run (sec.)} & \textbf{MCR} \\
\hline
MASS (500 iterations) & \phantom{0}55.34\phantom{00} & 0.187 (0.010) \\
SA-Sparse (500 iterations) & \phantom{0}44.70\phantom{00} & 0.301 (0.018) \\
SA-Sparse (5000 iterations) & 444.9275 & 0.237 (0.012) \\
\hline
\end{tabular*}
\end{table}

\begin{figure}[b]

\includegraphics{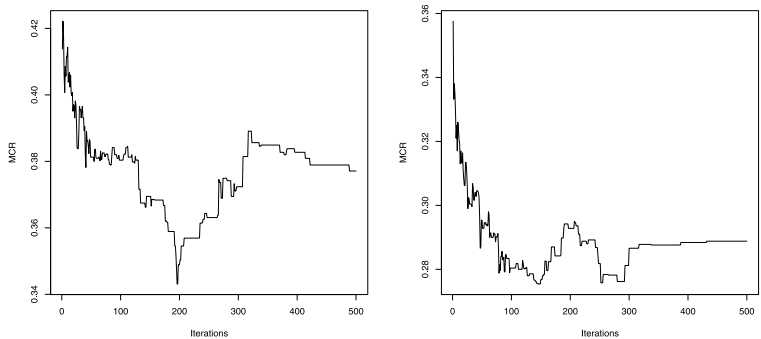}

\caption{Test MCR paths in Simulation I. Left panel: MFSS with
$\lambda=5$; right panel: MFSS with $\lambda=0$.}
\label{fig:overfit}
\end{figure}

Since the parameter space for MASS is large ($d\times L$ dimensional),
overfitting may be a potential problem. In our nonlinear simulation
study, we observed that when the model was very flexible, that is,
$\lambda$ was large, MFSS produced poor accuracy. Figure \ref
{fig:overfit} shows the average test MCR paths over 500 iterations in
the same setting as Simulation I except with $\lambda_0=1$. As is
displayed in the left panel of Figure \ref{fig:overfit}, when we
assign $\lambda=5$, which is a more flexible model than the true
model, the MCR decreases rapidly at the beginning, then it starts
increasing dramatically, and it levels off at some point of time. This
trend clearly indicates the presence of overfitting. When we used the
linear MASS (i.e., assigning $\lambda=0$), which was a less flexible
model than the true model, overfitting was less likely to occur. As we
can see from the right panel of Figure \ref{fig:overfit}, the MCR
decreases rapidly and levels off. In addition, we used a small value
for $p$ (the same as the true $p_0$). This made the model less
flexible. Presumably when $p \gg p_0$, overfitting may also become an issue.

\section{Discussion} \label{sec:6}

MASS was designed to implement a supervised learning classification method
with the flexibility to mimic either a variable selection or a variable
combination method. It does this by adaptively adjusting the sparsity of
the transformation matrix used to lower the dimensionality of the original
data space. We use a stochastic search procedure to address the very high
dimensional predictor space. Our simulation results suggest that this
approach can provide extremely competitive results relative to a large
range of classical and modern classification techniques in both linear
and nonlinear cases. We also examined
three different preliminary dimension reduction methods which appeared to
both increase prediction accuracy as well as improve computational
efficiency. MASS seems to converge quickly relative to other stochastic
approaches, which makes it feasible to be applied to large data sets.
The MASS method for dimensionality reduction could also be
generalized to the context of regression where the response is a
continuous variable. Further studies are planned for this setting.

\begin{appendix}\label{app:A}
\section{\texorpdfstring{Proof of Theorem
\protect\lowercase{\ref{mass.theorem}}}{Proof of Theorem 1}}
Suppose the theorem is not true. Then for some $\delta>0$ it must be the
case that as $n$ and $I$ converge to infinity, $\frac{1}{n} \|\tilde
\mathbf{Z}_n^{(I)} - \mathbf{Z}_0\|^2>\delta$ happens infinitely
often with a probability greater than
0. Without loss of generality, we can assume $\delta<\varepsilon$ where
$\varepsilon$ is defined in Assumption~\ref{var.select}. But since
MASS randomly searches the entire space of $\mathbf{Z}$, as
$I\rightarrow\infty$,
we are guaranteed at some stage to generate a candidate set of predictors,
$\tilde\mathbf{Z}_n^{(I)}$, such that $\frac{1}{n} \|\tilde\mathbf
{Z}_n^{(I)} -
\mathbf{Z}_0\|^2<\delta<\varepsilon$. The last point to prove is that
this candidate
set is selected by MASS and does not then get rejected at a later
iteration.

By Assumption~\ref{var.select}, there exists $\Omega$ with
$P(\Omega)=1$ such that for any $\omega\in\Omega$, for all
$n>N(\omega)$, $\tilde\mathbf{Z}_{n}(\omega)$ is guaranteed to be
selected since
$\frac{1}{n} \|\tilde\mathbf{Z}_{n}(\omega) - \mathbf{Z}_0\|
^2<\varepsilon$.\vspace*{-1pt} Once $\tilde
\mathbf{Z}_{n}(\omega)$ is selected, at each future iteration the
same set of
predictors will be presented to the variable selection method along with
other possible candidates. By Assumption~\ref{var.select}, the only way
that $\tilde\mathbf{Z}_{n}(\omega)$ would not be selected at the
next iteration
would be if an even better set of predictors was generated with squared
distance from the true predictors even lower. Hence, as $n$ and $I$
approach infinity, it must be the case that $\frac{1}{n} \|\tilde
\mathbf{Z}_n^{(I)} - \mathbf{Z}_0\|^2<\delta$. Thus, the theorem is proved.

\section{Deduction of nonlinear reduction} \label{app:B}

We write
$f_{j,k}(t)=\mathbf{b}(t)^T\theta_{j,k}=b_1(t)\theta_{j,k,1}+\cdots
+b_q(t)\theta_{j,k,q}$,
where $\mathbf{b}(t)$ is a NCS basis with $q$ degrees of freedom and
$\theta_{j,k}$ is the coefficient vector. We need to generate $\theta_{j,k}$
such that (\ref{smooth.constraint}) holds. First note that
$f''^2_{j,k}(t)=\theta^T_{j,k}\mathbf{b}''(t)\mathbf{b}''(t)^T\theta_{j,k}$.
Then the integral in (\ref{smooth.constraint}) becomes
\[
\int f''_{j,k}(t)^2\,dt \approx\frac{1}{T}\sum
^T_{l=1}f''_{j,k}(t_l)^2=\theta^T_{j,k}\Gamma\theta_{j,k},
\]
where $\Gamma=\frac{1}{T}\sum^T_{l=1}\mathbf{b}''(t_l)\mathbf{b}''(t_l)^T$ and
$t_1, \ldots, t_T$ represent a fine grid of time points. Applying the
singular value
decomposition, we can write $\Gamma=\mathbf{UDU}^T$,
where $\mathbf{U}$ is orthogonal and
$\mathbf{D}=\operatorname{diag}(d_1,\dots,d_{q-1},0)$. Note the $0$ in the
singular value decomposition comes from the slope term (set to zero when
you take the second derivative) since there are no intercept terms in the
basis function.

We further write
\[
\theta^T_{j,k}\Gamma\theta_{j,k}=\theta^T_{j,k}\mathbf{UDU}^T\theta_{j,k} =\theta^{*T}_{j,k}\mathbf{D}\theta^*_{j,k}
=\theta^{-T}_{j,k}\mathbf{D}^-\theta^-_{j,k},
\]
where $\theta^*_{j,k}=\mathbf{U}^T\theta_{j,k}$, $\theta^-_{j,k}$
is the
first $q-1$ elements of $\theta^*_{j,k}$, and
$\mathbf{D}^-=\operatorname{diag}(d_1,\break \dots,d_{q-1})$. Hence, we need to
constrain $\theta^{-T}_{j,k}\mathbf{D}^-\theta^-_{j,k} \le\lambda$.
This is easily\vspace*{1pt} achieved by first generating the $\theta_{j,k}$'s as
described in (\ref{eq:A1}) and (\ref{eq:A}), and computing the
corresponding $\theta^-_{j,k}$. We then reset $\theta^-_{j,k}$ via
\[
\theta^-_{j,k}\leftarrow
\theta^-_{j,k}\sqrt{\frac{\lambda}{\theta^{-T}_{j,k} \mathbf{D}^-
\theta^-_{j,k}}}
\]
and let $\theta_{jk}={\bf U}\theta^*_{j,k}={\bf U}(\theta
^-_{j,k},\theta^*_{j,k,q})^T$.

We write
\[
f_{j,k}(t)=\mathbf{b}(t)^T\mathbf{U}(\theta^-_{j,k},\theta
^*_{j,k,q})^T = (\mathbf{b}(t)^T\mathbf{U})\theta
^-_{j,k}+(\mathbf{b}(t)^T\mathbf{U})_q\theta^*_{j,k,q}.
\]
In particular, if $\lambda=0$, all elements in $\theta^-_{j,k}$ are
zero, in which case the integral is also zero and a linear fit is
produced, that is, $f_{j,k}(t)=(\mathbf{b}(t)^T\mathbf{U}
)_q\theta^*_{j,k,q}$.

Since $\theta^*_{j,k,q}$ indicates the slope term, standardizing
$f_{j,k}$ is equivalent to standardize all the $\theta^*_{j,k,q}$'s.
We let
\[
f^*_{j,k}(t)=(\mathbf{b}(t)^T\mathbf{U})_q\theta
^*_{j,k,q}=\upsilon t\theta^*_{j,k,q},
\]
where $\upsilon^2=\frac{1}{t}(\mathbf{b}(t)^T\mathbf{U}
)_q$. Standardizing $f_{j,k}$ involves setting
\[
\sum^d_{k=1}\upsilon^2\theta^{*2}_{j,k,q}=1.
\]
Hence, we reset $\theta^*_{j,k,q}$ by
\[
\theta^*_{j,k,q} \leftarrow\frac{\theta^*_{j,k,q}}{\upsilon_k\sqrt
{\sum^d_{k=1}\upsilon^2\theta^{*2}_{j,k,q}}}.
\]

This approach ensures that (\ref{smooth.constraint}) holds for all
candidate functions.
\end{appendix}
%

\printaddresses

\end{document}